\documentclass[letterpaper,12pt,final]{iopart}
\usepackage[utf8]{inputenc}
\usepackage{graphicx}
\usepackage{subfigure}
\usepackage{wrapfig}
\usepackage{lineno}

\usepackage{paralist}

\usepackage[colorlinks=true]{hyperref}
\hypersetup{pdftitle={}}

\begin{document}
\title{Particle Dance: Particle physics in the dance studio}
\author{K.~Nikolopoulos$^{1}$, M.~Pardalaki$^{2}$}
\address{$^{1}$ School of Physics and Astronomy, University of Birmingham, B15 2TT, United Kingdom}
\address{$^{2}$ Berdahi Company, Paris, France}
\ead{k.nikolopoulos@bham.ac.uk}
\date{March 2018}

\begin{abstract}
A workshop using dance to introduce particle physics concepts to young
children is presented. The workshop is realised in the dance studio,
the children assume complete ownership of the activity and dance
becomes the means to express ideas.  The embodiment of the physics
concepts facilitates knowledge assimilation, while empowering the
students with respect to science. Beyond the scientific and artistic
benefits of this workshop, this approach aspires to overcome the
barriers between art and science; and open new interdisciplinary
horizons for the students.
\end{abstract}

\submitto{\PED}

\section{Introduction}

The 2015 Nobel Prize in physics was awarded ``for the discovery of
neutrino oscillations, which shows that neutrinos have
mass''~\cite{nobel}. This was the starting point of a fruitful
collaboration between the authors, a deeper look at the respective
practices and exchange of ideas, that culminated in the ``Neutrino
Passoire'' performance~\cite{passoire}: following the elusive and
omnipresent neutrinos, from their birthplace in the Sun, travelling
through space and oscillating between flavours imperceptibly. In this
journey, the neutrinos transverse matter, the Earth, our bodies,
continuing unimpeded. This provides for the opportunity to explore the
idea that the human body is not a fortress as one might think; it is
rather perceived as a sort of passoire, the french word for colander,
letting neutrinos pass through without trauma or memory of the event
itself. From that point on, the performance naturally questions
notions particularly prevalent in the public discourse and leaves it
to the audience to provide answers.

The performance was presented for the first time at the University of
Birmingham Arts and Science Festival 2016, and in several venues
since. An extended version was presented at the Midlands Arts Centre
in March 2018, while the final version of the performance was
presented in Paris in October 2019. The overwhelmingly positive
feedback received from the audience compelled the authors to go beyond
the performance by taking particle physics in the dance studio,
allowing school students to approach the subject matter through an
experiential approach. This mirrors, to a large extent, the process
followed by the authors during the development of the performance.

Through the ``Particle Dance'' workshop, we aspire to make particle
physics more accessible, support students in developing
self-confidence in relation to science and research, and generate an
interest in science, as an inherently creative subject, that would
persist for a long time after the workshop.

\section{Workshop development} 
The ``Particle Dance'' workshop was developed, initially, as part of
the CREATIONS project, a Horizon 2020 support and coordination action
across eleven EU-member countries, aiming to develop art-based
creative approaches towards a more engaging science
classroom~\cite{CREATIONS}. The workshop design was informed by the
creative pedagogical features established within
CREATIONS~\cite{EXETER}, similarly to Ref.~\cite{Andrews:2018mjb}
which was developed in parallel.

The core idea is to bring particle physics in the dance studio. This
encourages informal learning, avoids any student pre-conceptions
regarding the science classroom, and allows for interdisciplinary
connections to be made. Furthermore, it allows students to appreciate
that dance is not only a means to express oneself in relation to
feelings and emotions, but also to convey ideas, even on topics that
are not usually perceived to be connected with dance. Thus, another
benefit of this approach, is the bridging of art and science subjects,
that are currently disconnected in the school curriculum. For this reason,
it is important that at the end of the workshop the students have a
complete artistic creation, a final product that they could perform
and discuss.

The workshop consists of two parts. In the first part, students learn
about fundamental particles that make up matter and mediate
interactions, as well as about the Higgs boson. At this stage,
students connect particles to dance through short choreographic
movements, proposed by themselves, inspired by the particle
properties, name, etc. In the second part students learn about
particle interactions, scattering, pair-production and annihilation,
and, in teams, produce a choreography of an interaction. They are
given complete responsibility both for the development of the
choreography and the choice of music.

During the workshop two main props are used:
\begin{inparaenum}[a)]
\item the ``subatomic
plushies''~\cite{particlezoo}, a cloth model for each particle, used to provide a visual anchor to the discussions; and 
\item a deck of particle trump cards, one per particle, showing the particle name and
basic properties~\cite{pavlidou}.
\end{inparaenum}

For the initial implementation phase of the workshop, reported here,
the main focus was on girls in Key Stage 3 (KS3), and more
specifically Year 8 (12-13 years old), in the British education system. Seven workshops
were delivered in schools in the West Midlands area with up to 16
students per workshop and a total of approximately 110 participating
students.

\section{Learning about the particles}
Some students in KS3 may be familiar with the idea that matter is made
out of atoms. However, it is very rare to find students that know the
details of the atom consisting of a dense nucleus made of protons and
neutrons, and electrons around them.  At the beginning of each
workshop, students sit in a circle on the floor and each one is
presented with a card, randomly drawn from a deck of particle trump
cards. These cards, originally developed for the ``particle physics in
primary schools'' workshop~\cite{pavlidou}, present each particle and
its properties in a simplified manner, appropriate for the age of the
students, and codify the interaction properties in terms of ``likes''
and ``dis-likes'' between the particles. At this stage 16 cards are
used, representing the matter particles, interaction carriers, and the
Higgs boson.

Subsequently, a model of the atom consisting of ``subatomic
plushies''~\cite{particlezoo} is presented and following discussion
about the atom consisting of a dense core of protons and neutron with
electrons orbiting around, the students are invited to ``open'' the
proton and the neutron to find that they are made up of up- and
down-quarks, along with gluons that keep them together.

\begin{figure}[h!]
\centering
\includegraphics[width=0.45\linewidth]{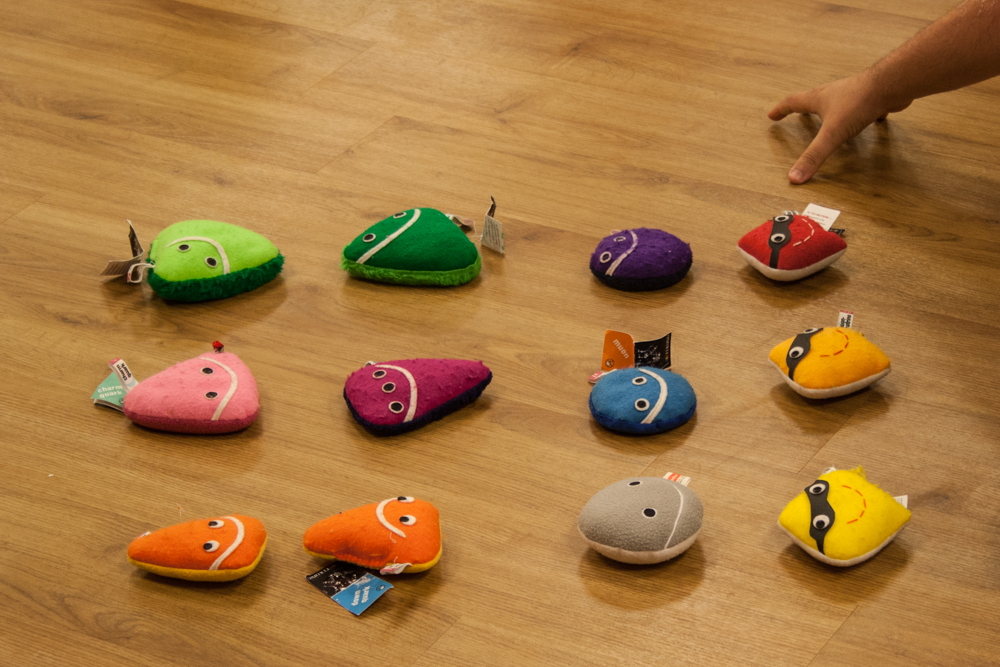}
\caption{Introducing the particle of the standard model using a ``plushie'' for each particle. Photo credit: Dimitra Spathara.\label{fig:plush}}
\end{figure}

Having completed this basic introduction with the atom, a discussion
begins between the physicist and the students. The physicist
introduces one-by-one the particles of the Standard Model, and the
students are asked one-by-one to read out loud the content of their
the trump card, as the particles are being presented and the ``plushies''
are arranged in familes, as shown in Fig.~\ref{fig:plush}. For each
particle, after the trump card is read, the physicist adds a funny or
interesting fact about the particles on top of the description read
out from the card, for example how they were discovered, how they got
their name, etc.  The use of these trump cards was found to be very
effective in engaging the students, as each of them has ``their own
particle'', and provides a natural opportunity for each student to add
something to the discussion. Thus, it also acts as an ice-breaker, and
allows for active participation in the ensuing discussions.

Having introduced the particles of the Standard Model, a discussion
which lasts for about 25 minutes including questions and answers, the
warm-up begins. Each student holds a cloth model of a
particle. Students start walking in the dance studio, changing
direction freely, engaging in eye contact with their classmates. When
two students simultaneously look at each other, they exchange their
particles by throwing them to one another, an activity that links to
particle interactions in the microcosm. Progressively, walking becomes
faster, almost running. Students are asked to move through space
constantly being aware of the presence of the others. Eventually this
includes jumps, change of levels, and balance exercises. The ``plushie
particles'' in dance exercises provide concrete support and a movement
reference to the student who is warming up while playing. At the same
time they provide an excuse for the students to interact with one
another. The warm up ends in a circle with some stretching and
exercises performed in tandem.

\begin{figure}[h!]
\centering
\subfigure[Up~quark]{\includegraphics[height=4cm]{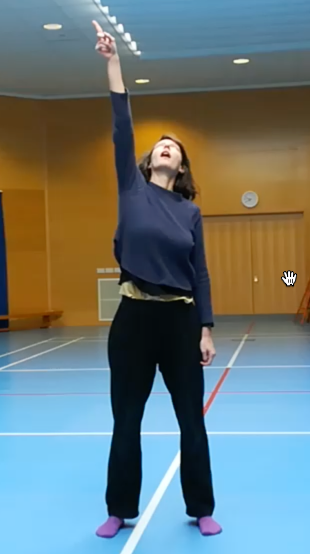}}
\subfigure[Down~quark]{\includegraphics[height=4cm]{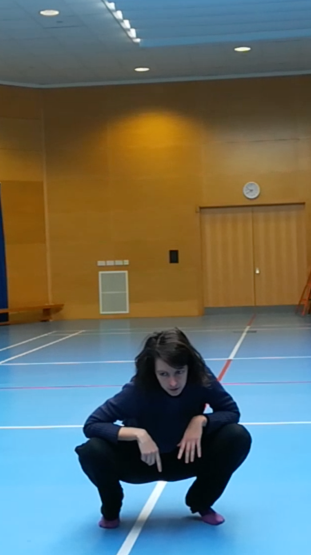}}\\
\subfigure[Charm~quark]{\includegraphics[height=4cm]{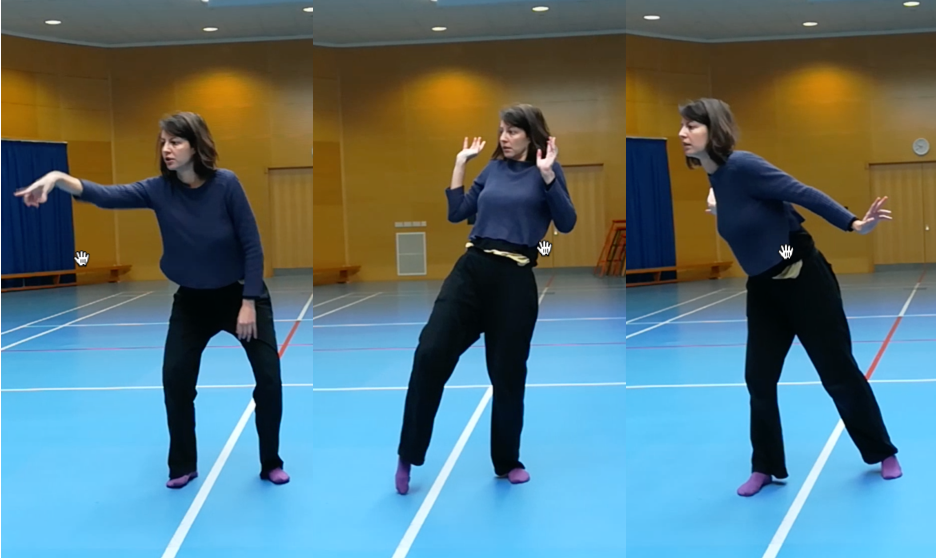}}
\subfigure[Strange~quark]{\includegraphics[height=4cm]{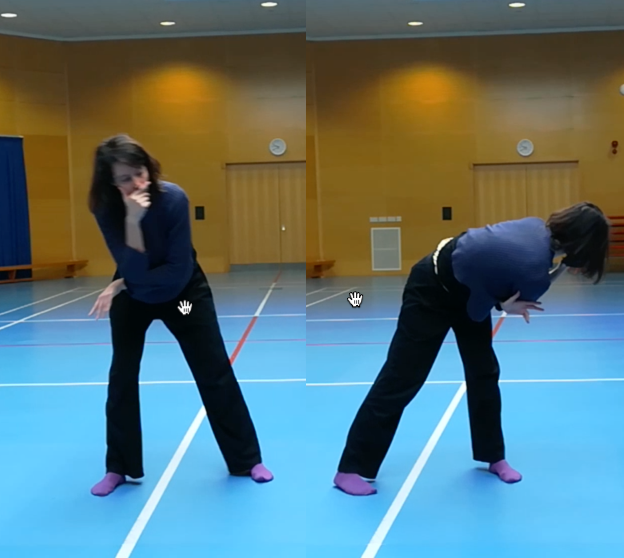}}\\
\subfigure[Top~quark]{\includegraphics[height=4cm]{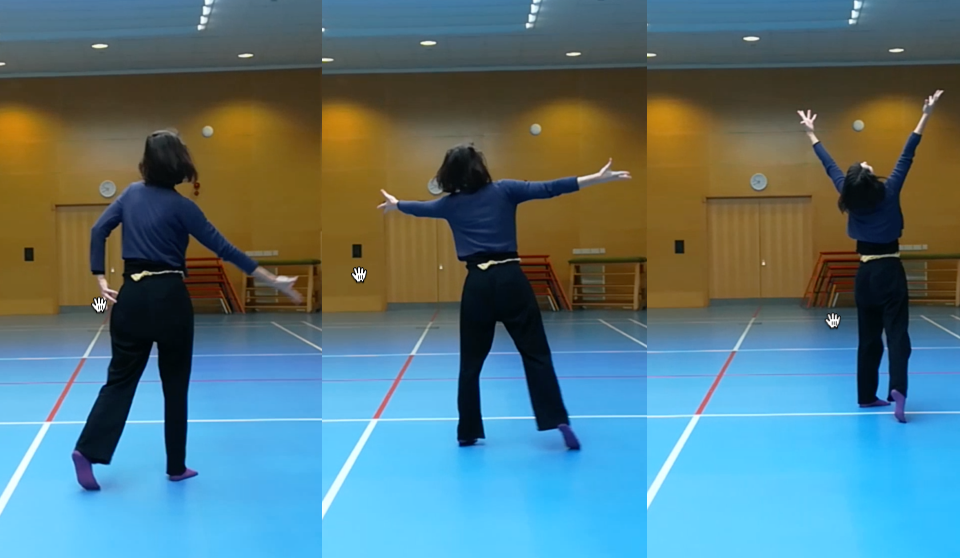}}
\subfigure[Beauty~quark]{\includegraphics[height=4cm]{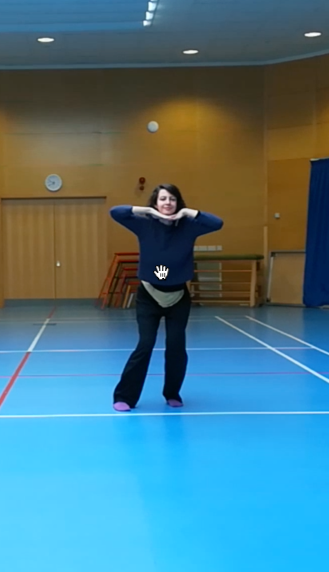}}
\caption{Example choreography for the quarks by Mairi Pardalaki. Extracted from video by Daniela \'{A}mbar Gayoso-Miranda.\label{fig:choreography}}
\end{figure}

Following the warm up, the first exercise begins. Each particle is
associated with a dance move, inspired by its name, its properties, or
anything else that may have drawn the interest. Initially, the dancer
provides a set of moves associated to the particles, and subsequently
each of the students in turn adds a move for the particle in their
card. The students are asked to guess the particle, when the dance
teacher is showing a move. The students succeed in guessing most of
the particle moves, as they have read the particle identity cards in
the beginning of the workshop. The teachers use simple, almost
theatrical moves to embody the particles, and the students are
encouraged to do the same. An example choreography is given in
Fig.~\ref{fig:choreography}. When a student is hesitant to propose a
movement, the teachers take this unconsciously produced hesitant
movement and put it into the choreography. The choreography is
repeated with the dance teacher naming the particles at the beginning
of every movement. This way the students assimilate the name, together
with a body movement.

In this exercise, the primary role of the music is to support
the developping choreography. However, the presence of the musician
and the interpretation of the evolving choreography through improvised music,
adds another layer of dialogue, between the dancers and the
musician. This way, the musician, seeing the evolving choreograph, can
add or propose surprise music elements that enrich, and potentially
guide, the process. This way the imaginative space created between
science, dance, and music is expanded.

\section{Learning about anti-matter and particle interactions}
In the second part, the students are introduced to the idea of
anti-matter. The physicist demistifies anti-matter: there is nothing
mystical or apocryphal about it and it is routinely produced in cosmic
ray collisions in the atmosphere and high energy particle
colliders. The key element of particle annihilation is presented. This
is very exciting for the students, they show visible interest in the
discussion, and they ask several questions trying to grasp the nature
of these anti-matter particles, and the process of annihilation where
photons appear where matter and the, respective, anti-matter particles
meet. At this point also the notion of pair-production of particle is
introduced, and this, naturally, leads to the possibility of
scattering between particles.

\begin{figure}[h!]
\centering
\subfigure[\label{fig:chorA}]{\includegraphics[width=0.45\linewidth]{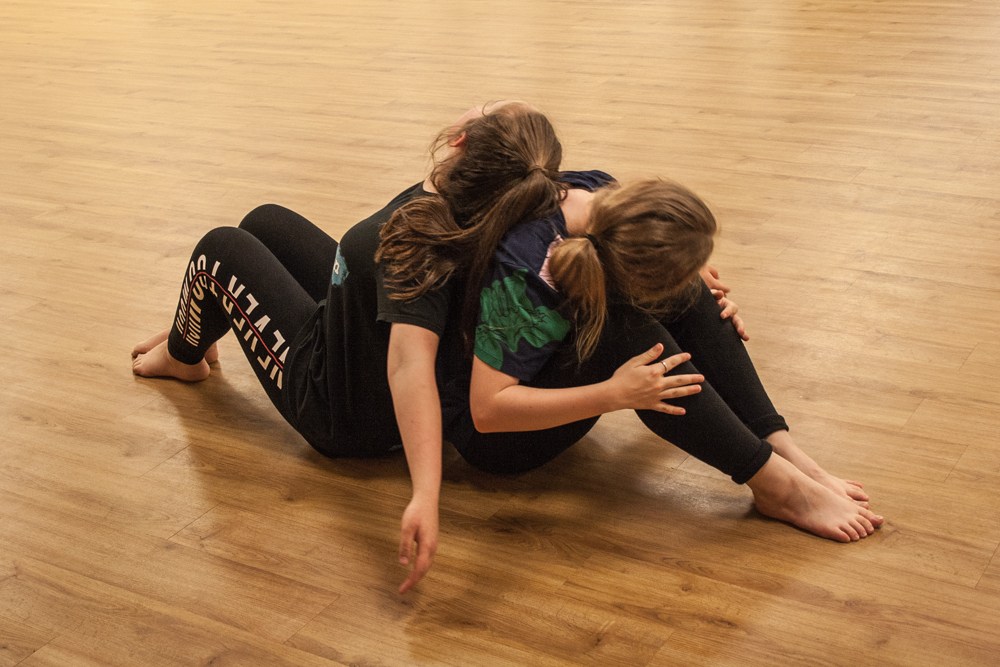}}
\subfigure[\label{fig:chorB}]{\includegraphics[width=0.45\linewidth]{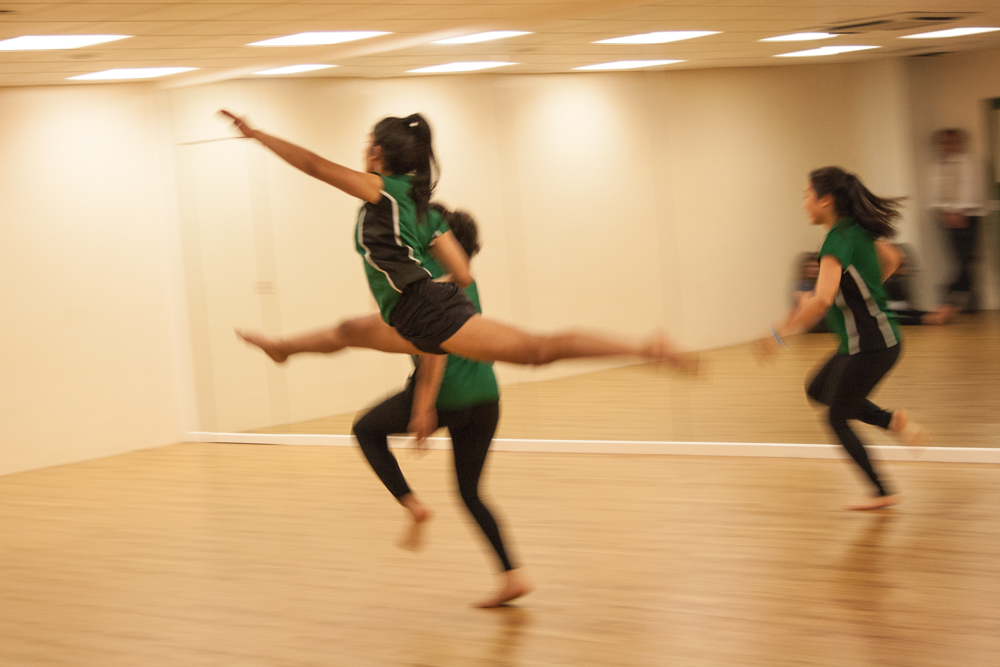}}
\caption{Choreographed interactions by teams of students. Photo credit: Dimitra Spathara.\label{fig:chor}}
\end{figure}

Subsequently, the students are split in teams, with each team tasked
to produce a choreography for a given interaction.  Particle
scattering is choreographed by a team of three students (two particles
and a force carrier), the annihilation and subsequent pair production
by a team of five students (particle and anti-particle in the initial
state, force carrier, particle and anti-particle in the final state). A
team of seven students may produce a choreography of the atom (three
quarks, two gluons, one electron, and a photon). Initially, the team should
decide which particles they are going to use. A requirement is that at least one of the moves developed
in the first part should be retained. It is interesting to observe 
how the students humanise the inanimate particles by giving them facial expressions, funny interactions, and by using dance techniques they are
familiar with, for example ballet, jazz, or modern.
First, they imagine the interaction and then translate it and
enrich it through dance, see Fig.~\ref{fig:chor}.

It is crucial that the students not only decide on the particles to
use and the choreography, but also the music itself. Each team, once
they have an initial plan for the choreography and the music,
discusses directly with the musician on the music they would like to
use for their performance. This has been proven to be a key point for
the students to take full ownership of the final product. In most
cases, it was the first time that they have this kind of freedom
during a dance class.

\section{Particle Dance and open questions in particle physics}
At the end of the session, both parts are presented as a whole, the
``Particle Dance'', and the choreography is completed and the students
feel particularly excited about performing.

Finally, the students sit again in a circle on the floor and a
discussion takes place. This opportunity is used to introduce elements
of open questions. The students are confronted with the idea that the
Standard Model particles, all those discussed during the workshop,
make up only about 4.6\% of the matter-energy content of the universe,
and that the majority is attributed to Dark Matter and Dark
Energy. The reaction of the students to this varies, most of them are
excited and interested, some are disappointed. One student actually
got upset, almost angry, not being able to accept the relevation!

\section{Evaluation}
At the end of each workshop a round table takes place were
students are invited to provide feedback on the session, and
comment on things that they did or did not enjoy about it, and
suggestions for improvements. At the same time feedback is requested
from the teachers accompanying the students to the workshop. Overall,
the received feedback has been particularly encouraging.

Based on the above the workshop seems to hold promise in achieving its
aspiration to make particle physics more accessible, help students
build confidence in themselves in relation to recent scientific
developments, and -- eventually -- create images that the students
will carry in their mind long after the workshop has been completed.

The students were found to be particularly excited about the creative
aspects of the event and how they developed their own dance.  Prior to
the event the students had little idea of what it would involve and
they expressed surprise at how physics and dance came together so
naturally They found the event different and unusual.  The use of
``plushies'' was positively noted as it made easier to visualise certain
aspects of the particle properties.  The collaborative aspects, and
working as a group, were positively noted.

Obtaining a final ``end product'' combining both parts of the workshop
in one choreography, proved a powerful concept appreciated by the
students.  It was though that interleaving physics with dance parts
was beneficial for their understanding.

One of the students mentioned that they appreciated how the dancers
would take their suggested movements in the first part, and turn them
in actual dance moves. The aspect of having live music, and in
particularly chosing the music on their own proved extremely powerful,
to the extend that it was thought to be the cornerstone in the process
of the students taking ownership of the choreography. Usually, in
dance lessons the students are given a song or a piece of music to
dance to, and the approached used in this workshop was ground-breaking
for them.

Regarding the learning outcomes, students indicated that they learnt
what is inside the atom. Most of them thought that the atom is indeed
the smallest division of matter, and they were particularly excited to
find out how many different particles there are and that they can be
created or disappear through their interactions. ``I always thought it
was just one atom and there were just electrons going around it. But
actually there is so much more to it.'' one student commented.

The interdisciplinarity aspect was highlighted. Students, were
very keen to respond to the questions at the end of the workshop, a
striking difference with their relunctance at the beginning of the
workshop. They definitely had increased confidence in relation to both
expressing their opinion, likes and disklikes of the workshop, but
also on discussing the bits and pieces they have learned during the
day.
In a scientific context this is used to explain that particle
physicists work collaboratively, nowadays in very large teams from
different countries, and that collaboration, and the ability to
communicate and exchange ideas is crucial for progress.

\section{Summary}
With the development of the ``Particle Dance'' workshop we aspire to
bring particle physics in the dance studio, and to stimulate the students' 
curiosity towards particle physics and science in
general. The workshop builds on the CREATIONS creative pedagogical
features Integral part of the workshop is the introduction of novel
for the students particle physics concepts. The evaluation of the
pilot phase is encouraging regarding the effectiveness of the
workshop. Future steps involve focused implementation and expansion to
different ages to allow for larger classrooms, and the training of
science-art teachers.

\ack 
This work has been supported by the European Commission as part
of the ``CREATIONS - Developing an engaging science class room''
project (CREATIONS-665917) and by the West Midlands Branch of the
Institute of Physics.  The participation of dancer Fanny Travaglino,
and musicians Alexandros Mentis and Katerina Fotinaki, in the
workshops, along with their insightful comments is thankfully
acknowledged.

\section*{References}
\bibliographystyle{ieeetr}
\bibliography{references}
\end{document}